\documentclass[%
twocolumn,
groupedaddress,
nofootinbib,
aps,
prx,
]{revtex4-1}

\usepackage{hyperref,url}

\newcommand{\beq}{\begin{equation}}
\newcommand{\eeq}{\end{equation}}
\newcommand{\ben}{\begin{eqnarray}}
\newcommand{\een}{\end{eqnarray}}

\begin{document}

\title{Feasibility of Net Energy Gain in Kinematic Nuclear Fusion
Devices}

\author{E.V. Tsiper}

\affiliation{e.tsiper@yahoo.com}

\date{May 21, 2020}

\begin{abstract} A design principle is suggested to overcome obstacles
that prevent positive net energy output in nuclear fusion devices
based on electrostatically accelerated ions. Since Coulomb scattering
cross-section dwarfs that of nuclear fusion, the focus is on
re-capturing energy of elastically scattered ions before the energy is
lost to heat. Device configuration to achieve efficient energy
re-capturing is proposed and a favorable estimate of net energy gain is
obtained. \end{abstract}

\maketitle

\section{Brief Problem Overview}

Nuclear fusion reaction occurs when two ions of certain kind hit each
other at energies large enough for them to overcome Coulomb repulsion
and approach each other at distances of the order of $10^{-14}$m,
about $10^4$ times smaller than the size of an atom. Consequently, the
kinetic energy required is $10^4$ times larger than typical chemical
energies: tens or hundreds keV, or, equivalently, hundreds or millions
K in the temperature scale.

A majority of fusion experiments attempt to heat plasma to the
required temperatures so as to allow random ion collisions to cause
fusion reactions. The high temperatures involved require confinement
of the plasma, either magnetic or electrostatic, in order to spare the
apparatus from the hot plasma inside. Confining and maintaining hot
plasma is a formidable task that has yet to result in a controlled
sustainable fusion reaction.

An alternative approach is to accelerate the ions by means of electric
field, which only requires modest voltages of tens or hundreds kV.
However, certain fundamental obstacles reviewed below are believed to
preclude net energy gain in such a set up. ``Highly nonthermal
systems, like the colliding beam reactor proposed by Rostoker et
al.~would relax to local thermal equilibrium before a significant
amount of fusion power could be produced.''\cite{nevins,nevins1995}
Here I propose a way to overcome these fundamental obstacles.

The first example of why kinematic approaches do not work is usually
to consider an energetic ion beam hitting a solid target. Since the
fusion cross-section is so small compared to the squared distances
between the atoms in the target, an average ion will have to traverse
great many atomic layers before it has a chance of hitting a
nucleus. Such an ion will be stopped at a much shorter distance by
electrons in the target.

Even in configurations with no electrons present, a fast ion has a
much greater chance of hitting another ion close enough to scatter
elastically than to hit it dead-on to cause fusion. The ions that
scatter elastically redistribute their kinetic energies between them,
causing cascading process that quickly leads to thermalization --- the
loss of the initial kinetic energy to heat.  There are only two
possible outcomes of thermalization: either the heat leads to the
overall temperature sufficient for sustaining fusion, bringing the
device to the class of the plasma confinement devices (not discussed
here), or (ii) the temperatures are lower than fusion temperatures, in
which case the energy lost to heat is unrecoverable fully due to the
laws of thermodynamics, thereby precluding net energy gain.  In other
words, unless a very hot plasma is formed, much more energy has to be
spent on accelerating the ions that are unsuccessful at fusion than
any gain produced by the very few ones that are successful.

The second example is a fusor device \cite{farnsworth} --- the simplest
device that does achieve fusion reaction by means of electrostatic
potentials, albeit no net positive gain has been demonstrated so far.

It is crucial to underscore that the task of achieving fusion is not
hard; a fusor is a simple enough device to be built at home or in a
garage setting. The difficult task is to make a
sustainable fusion reaction that can feed itself through net energy
gain. Energy losses in a typical fusor device are five orders of
magnitude larger than the fusion power produced \cite{hedditch}.

A typical fusor device does not fall into the class of kinetic fusion
approaches reviewed here; it is rather described as an Inertial
Electrostatic Confinement (IEC) device --- the one that still employs
hot plasma, shielded from the outside apparatus by electrostatic,
rather than magnetic, field (see \cite{bhattacharjee} and
bibliography therein). A well-known critique of IEC devices is
contained in \cite{rider}. The reason for fusors being IECs
is a fundamental channel for energy thermalization, similar to that
described above. The ions in the fusor device are accelerated by
electrostatic bias when they fly from the outer wall towards the
center. In the center the average kinetic energy of ions is large
enough to undergo fusion. Two cold ions accelerated towards the center
from the outer wall have the same energy when they reach the
center. They have a chance to hit each other and cause fusion, but
they also have a much greater chance to scatter elastically and
re-distribute their energy between them.

When two particles of the same energy scatter elastically they may
still re-distribute their energies.  This can be visualized on a
billiard table: it is possible, for instance, for two identical
billiard balls with the same speed to collide in such a manner that
one of them stops and the other flies away with twice the energy. The
process is the exact reverse of a billiard ball hitting a stationary
billiard ball.  In summary, the energy outcome of an elastic collision
depends on timing and mutual trajectories even for identical particles
with identical initial kinetic energies. Energy re-distribution due to
Coulomb scattering leads to thermalization.

The same fundamental obstacle applies to other configurations that
attempt to achieve sustainable fusion via accelerated ion beams, such
as, for instance, in accelerated beam fusion reactors (ABFR). In
addition to the usually-quoted problem with beam de-focusing due to
internal electrostatic pressure, the same process causes many more
ions to be elastically scattered away, carrying their energy with
them, than the few that cause fusion, precluding net energy gain.

Here I discuss a device configuration, which circumvents
the above fundamental obstacle by allowing efficient recovery of the
energy carried away by the elastically scattered ions \cite{pending}.

\section{Preventing Energy Redistribution of Colliding Ions}

Energy re-distribution is absent for two particles of the same mass
and same but exactly opposite velocities. Elastic scattering in such
head-on collisions result in both particles changing the direction of
their flight but not their energies. Scattered particles fly away in
the opposite directions that can be at any angle to their initial
velocities, but their energies remain equal to their initial
energies. This property can be used, which is the focus of this work,
to help recover the energies of the scattered ions.

Deviation of the collision angle from $180^\circ$ quickly breaks down
this property, allowing for energy redistribution, thereby making the
task of recovering the energy of the scattered particles much harder,
if not impossible.  Employing ion species with different masses, such
as D--T or p--$^{11}$B, also leads to energy re-distribution.  This
narrows down the present approach essentially to D--D, T--T or the
aneutronic $^3$He--$^3$He fusion reactions.

\section{Worked Example}

As an illustration of the concept, consider a vacuum-evacuated chamber
containing two cold ion injection openings of small angular size
$2\alpha$ placed opposite to each other and a coaxial ring-shaped
accelerating electrode in the center, negatively-biased with respect
to the chamber.  If necessary, the paths of the opposite ion beams can
be controlled by small deflecting magnetic fields.  If the paths are
made to collide head-on near the center of the accelerating electrode,
the ions will have a chance to undergo a fusion reaction, provided the
bias voltage is large enough.

The majority of the ions will not scatter or undergo fusion, but will
de-celerate as they reach the opposite side of the chamber, returning
their energy back to the cirquit.  In the simplest basic design the
path of these un-scattered ions can be made to hit the opposite cold
ion injector, in which case these ions can be either collected or
allowed to return back to the chamber and be accelerated again for
another try.

The great majority of the rest of the ions will be scattered
elastically in near-head-on collisions.  These ions will have a very
narrow energy distribution and a certain distribution over scattering
angles.  These ions too will be decelerated by the positive electrode
and return the bulk of their energy to the circuit.  Now, these
scattered ions will have to be collected and removed from the chamber
by the evacuation system (the positive electrode will have to be made
of a mesh).

It is imperative to not allow the ions scattered at angles greater
than some value (say, of the order of $\alpha$) to return back to the
chamber, as the next scattering event will not be head-on and will
cause energy re-distribution.

Thus, ignoring fusion reactions for the moment, the steady state of
the device will have a highly non-thermal spatial and velocity
distribution strongly enhancing same-mass same-energy head-on
collisions near the center of the chamber.

The concept design resembles an usual fusor device.  However, the back
energy transfer is made possible by restricting the majority of the
collisions to be head-on collisions of same-mass, same-energy ions,
whereby the energies of the scattered particles are known exactly to
within narrow tolerances.  Prior-art fusor designs lack this critical
ability to retrieve the post-scattered ion energy fully and to avoid
thermalization.  Therefore, the operation of the present device
differs significantly from that of the prior-art fusor, with
high-temperature plasma in the central area replaced with a narrow,
highly non-thermal phase space distribution.

Additionally, prior-art fusor devices suffer from energy losses due to
the hot ions striking the central accelerating
electrode \cite{hedditch}, whereas the current design eliminates this
second critical energy loss channel, because the 1D path of the ions
does not cross the ring-shaped accelerating electrode.

The operation of the present device may have some semblance with the
``star mode'' of the usual fusor device, but with the “star” being
only two-pronged.

\subsection{Deviation from Head-on Collisions}

Whereas ideally the post-scattered ions all have the same energy,
finite injector size and other technological imperfections lead to a
(narrow) distribution of energies.  The energy recovery electrode
surrounding the reaction chamber must be under-biased in order to
disallow return of low-energy post-scattered ions to the reaction
chamber.

Deviation from the head-on collision by an angle $\delta$ leads to the
post-scattered ions having energy slightly above and slightly below
the initial energy. The worst-case scenario (the largest energy
deviation $dE$) occurs for the ions scattered at right angle, in which
case the velocity component normal to the main axis is either added or
subtracted from the post-scattered velocity.  In this case the energy
excess/deficiency is $dE(90^\circ)=2E_0\sin(\delta)$, where $E_0$ is
the accelerating electrode bias.  Absent additional beam focusing
elements, the angle $\delta$ can be estimated as $\delta\sim\alpha$,
half the angular size of the ion injection opening, as seen from the
center of the device.  The energy deficiency $dE$ may not be fully
recovered and thus contributes to losses.

Since the great majority of Coulomb scattering events occur to small
scattering angles, the areas of the energy recovery electrode near the
axis may be biased closer to $E_0$ in order to limit the losses. The
energy recovery electrode may be made segmented to achieve this goal.

\subsection{Net Energy Gain Feasibility Estimate}

Formally, the long-range nature of Coulomb force in vacuum makes every
ion scatter, albeit to a small angle. For the technical purpose of
this description I call ``un-scattered'' the ions that, upon passing
the reaction zone, scatter at angles less than $\alpha$.

In the simplest concept design these ions are permitted to be
reflected back and to accelerate again towards the center making
multiple attempts at the fusion reaction, until they scatter
away from the head-on collision trajectory (most likely), leading to
some energy loss, or undergo fusion.

I estimate the gain/loss balance assuming a 5 mm ion injection opening
diameter in a 30 cm diameter reaction chamber and the bias voltage of
500 kV for D--D reaction.  Thus, ``un-scattered'' ions are those ions
that scatter at angles less than $\alpha\approx1^\circ$.

The ions that scatter at angles greater than $\alpha$ have the Coulomb
scattering cross-section

\beq
\sigma_C=\frac{\pi}{16}\left(\frac{e^2\cot\alpha/2}{4\pi\epsilon_0E_0}\right)^2=235\;\text{barns}
\eeq
 This large number is to be compared against the DD fusion
cross-section $\sigma_F$ of only 0.19 barn --- the dramatic mismatch,
which exemplifies the hurdles of the kinematic fusion devices, and
which the present design is attempting to overcome.

In other words, for every ion pair undergoing fusion reaction, the
number $\sigma_C/\sigma_F=1200$ pairs are scattered elastically
away from the head-on trajectory without undergoing fusion.  The
kinetic energy of these ions needs to be recovered as fully as
possible to achieve net energy gain.

On the positive side of the net energy balance is the energy
$E_F=3.61$ MeV released by a successful fusion reaction.

Now I estimate the residual energy loss $dE$ per ion due to
deviations from head-on collision and the need to under-bias the
energy recovery electrode by this amount, as mentioned.  A very crude
estimate can use $dE(90^\circ)\approx 2E_0\sin(\alpha)$, which in our
case amounts to abour 3.3\% of $E_0$.

However, this energy loss can be reduced further by realizing that the
greatest fraction of the ions that do scatter are scattered to small
angles for which $dE$ is much smaller.  The energy of ions scattered
at an arbitrary angle $\chi$ lies between $E_0-dE(\chi)$ and
$E_0+dE(\chi)$ where

\beq
  dE(\chi) = 2\sin\alpha\;\sin\chi\;E_0.
\eeq
The highest energy defect, $dE(90^\circ)$, is observed for the ions
scattered at the right angle, but the fraction of these ions is
very small compared to the fraction of the ions scattered to small
angles. For the ions scattered at $\chi=1^\circ$ the energy defect is
only about 0.06\% of $E_0$.

The weighted average $\overline{dE}$ over all scattering angles
$\chi>\alpha$ is about 0.13\% of $E_0$, assuming the energy recovery
electrode is made segmented, each segment at angle $\chi$ being biased
with the bias defect of $dE(\chi)$.

Assuming efficiency $\eta$ of the fusion energy recovery, the energy
balance has, on the gain side, $\eta\sigma_F(E_F+2E_0)$ per ion pair
vs.~$0.0013 E_0\sigma_C$ per ion on the loss side.  The Gain/Loss
ratio is, therefore,

\beq
 G/L=\frac{\eta\sigma_F(E_F+2E_0)}{2\sigma_C\overline{dE}}\approx2.8\eta,
\eeq
 attesting to technical feasibility of the overall scheme.

Granted, it is still not a trivial task to achieve $G/L>1$ in a
practical device; however, the proposed approach replaces the
fundamental obstacle with an engineering challenge.

Higher practical $\eta$ values are facilitated for D--D reaction by
the fact that 63\% of the fusion yield is carried away by charged
particles (vs.~only 20\% for D--T), allowing direct energy conversion.

$G/L$ grows with $E_0$ due to decreasing $\sigma_C$.  The 500 kV bias
seems to be within the bounds imposed by electrical vacuum breakdown
\cite{slivkov}, as is the electric field $\sim2\cdot10^7$ V/m at the
central electrode (6cm diameter assumed).  If necessary, the field
parameters can be relaxed by scaling up the linear dimensions of the
device.

\subsection{Beam Defocusing Estimate}

\subsubsection{Defocusing due to the initial ion temperature.}

Assuming the cold ion injector at temperature $T$, the normal
component of the thermal motion of ions is of the order
$v_n\sim\sqrt{k_BT/m}$, where $m$ is the ion mass. In the simplest
concept design under consideration, absent additional beam focusing
devices, this velocity component contributes to beam defocusing and
consequent deviation from head-on collision via the time-of-flight for
the ions. Depending on the device parameters and dimensions, the cold
ion injector may need to be kept at cryogenic temperatures to limit
thermal defocusing.

The time-of-flight to the center is

\beq
 t\approx\frac{R_2}{v_0} \times\frac{\pi}{2}\sqrt{\frac{R_2}{R_1}},
\eeq
 where $R_2$ is the radius of the reaction chamber, $R_1$ is the radius
of the accelerating electrode, and $v_0=\sqrt{2E_0/m}$ is the hot ion
velocity.  For $R_1=3$cm this leads to thermal defocusing of less than
0.1 mm for injector at room temperature (0.2 mm contributed to the
beam diameter).

\subsubsection{Internal Coulomb defocusing}

The usual critique of accelerated beam fusion reactors conclude that
the beam densities necessary to achieve certain energy output lead to
beam self-defocusing because of internal Coulomb repulsion
\cite{liu,sands}. Here I do not make any claim of large energy output,
which may be limited by this and other considerations. The goal here
is only to design a device with net-positive energy output, albeit
possibly small.

On the other hand, Coulomb defocusing limitations can be alleviated or
relaxed by including beam focusing elements (conveniently made easier
by the 1-dimensional sptial distribution) or employing more involved
designs, including e.g.~separating ion injectors and ion collectors
via deflecting magnetic fields, replacing fusor design with beam
storage rings as in Ruggiero \cite{ruggiero}, employing a single
8-shaped self-crossing ring etc.

Any such design needs to implement the basic design principle to (i)
strongly enhance same-energy head-on collisions in the reaction zone,
(ii) efficiently evacuate the ions scattered to angles inconsistent
with the 1D phase space distribution maintained and (iii) to
efficiently collect the kinetic energy of such post-scattered ions and
return it back to the electric circuit before thermalization occurs.

I would like to thank V.M.~Belyaev for helpful discussions.


\begin{thebibliography}{11}

\bibitem{nevins} W.M.~Nevins et al, Science 281, 307, 1998.

\bibitem{nevins1995} W.M.~Nevins, Phys. Plasmas 2, 3804, 1995.

\bibitem{farnsworth} P.T.~Farnsworth, US Patent 3,258,402, 1966.

\bibitem{hedditch} J.H.~Hedditch, R.~Bowden-Reid, J.~Khachan, Phys.
Plasmas 22, 102705, 2015.

\bibitem{bhattacharjee} D.~Bhattacharjee, arXiv:2002.05941 [physics.plasm-ph], 2020.

\bibitem{rider} T.H.~Rider, MIT Thesis, 1994.

\bibitem{pending} Patent pending.

\bibitem{slivkov} I.N. Slivkov, V.I. Mikhailov et al., {\em  Electrical Breakdown and Discharge in Vacuum}, Atomizdat, Moscow, 1966.

\bibitem{liu} K.-F.~Liu and A.W.~Chao, Nucl. Fusion 57, 084002, 2017.

\bibitem{sands} M.~Sands, Conf. Proc. C 6906161 257, 1969.

\bibitem{ruggiero} A.G. Ruggiero, 2000 Proc. of ICONE 8, 8th
Int. Conf. on Nuclear Engineering (Baltimore, MD, USA, 2–6 April
2000).

\end{thebibliography}
\end{document}